\documentclass[12pt,tightenlines]{revtex4-1}   
\usepackage{graphicx}
\usepackage{hyperref}
\hypersetup{
  pdfauthor={Florian Sedlmeir, Harald G. L. Schwefel},
  pdftitle={High-Q MgF2 whispering gallery mode resonators for refractometric sensing in aqueous environment},
  pdfkeywords={Physics, Optics, Whispering Gallery Mode Resonators, sensing},
  urlcolor=blue,
}
\usepackage{amsmath}
\usepackage{amssymb}
\usepackage{siunitx}
\usepackage[utf8]{inputenc}

\usepackage[margin=1cm,labelfont=bf,font=small,justification=centerlast]{caption}

\newcommand{\mgf} {MgF$_2$ }

\begin{document}

\title{High-Q \mgf whispering gallery mode resonators for refractometric sensing in aqueous environment}

\author{Florian Sedlmeir$^{1,2,3,\dagger}$}\email{florian.sedlmeir@mpl.mpg.de} 

\author{Richard Zeltner$^{1,2,\dagger}$, Gerd Leuchs$^{1,2}$, and Harald G.L. Schwefel$^{1,2}$}

\affiliation{$^1$ 
Max Planck Institute for the Science of Light, G\"unther-Scharowsky-Straße 1/Building 24,
91058 Erlangen, Germany}
\affiliation{$^2$ Friedrich-Alexander-Universit\"at Erlangen-N\"urnberg (FAU), Department of Physics, Institute for Optics, Information and Photonics, Staudtstr.7/B2, 91058 Erlangen, Germany}
\affiliation{$^3$SAOT, School in Advanced Optical Technologies, Paul-Gordan-Str. 6,
91052 Erlangen, Germany \\ \\ $^\dagger$ authors contributed equally }



\begin{abstract}
We present our experiments on refractometric sensing with ultrahigh-$Q$, crystalline, birefringent magnesium fluoride (MgF$_2$) whispering gallery mode resonators. The difference to fused silica which is most commonly used for sensing experiments is the small refractive index of \mgf which is very close to that of water. Compared to fused silica this leads to more than $50\%$ longer evanescent fields and a $4.25$ times larger sensitivity. Moreover the birefringence amplifies the sensitivity difference between TM and TE type modes which will enhance sensing experiments based on difference frequency measurements. We estimate the performance of our resonators and compare them with fused silica theoretically and present experimental data showing the interferometrically measured evanescent decay and the sensitivity of mm-sized \mgf whispering gallery mode resonators immersed in water. They show reasonable agreement with the developed theory. Furthermore, we observe stable $Q$ factors in water well above $1 \times 10^8$.
\end{abstract}
\maketitle

\section{Introduction}

Whispering gallery mode (WGM) resonators have been established as a powerful tool for various sensing application over the last decade. Such dielectric resonators confine light via total internal reflection close to their surface causing an evanescent field leaking out into the environment. The resulting high-$Q$ resonances are thus sensitive to parameter changes of the resonator itself as well as of the surrounding. The applications range from refractometric sensing \cite {hanumegowda_refractometric_2005,zamora_refractometric_2007}, temperature \cite{guan_temperature_2006} and pressure sensing  \cite {ioppolo_micro-optical_2008, weigel12} to the detection of nearly any kind of biological matter like viruses, microorganisms and proteins \cite{vollmer_single_2008, ren_high-q_2007, arnold_shift_2003} even down to the single molecule level \cite {dantham_label-free_2013,vollmer_whispering-gallery-mode_2008}.

The most common types of WGM resonators used for sensing are fused silica microspheres \cite{arnold_shift_2003}, silica microtoroids \cite{hunt_bioconjugation_2010} and polystyrene microbeads \cite{lutti_monolithic_2008}. These devices share isotropic optical properties, high refractive indices compared to water and their amorphous nature. In this study we introduce and investigate for the first time ultrahigh-Q single crystal WGM resonators for sensing purposes. We chose magnesium fluoride (MgF$_2$) as the resonator material because its refractive index is very close to that of water and it is thus expected to show a stronger sensitivity towards the surrounding. 
Moreover it is slightly anisotropic and therefore the difference in responsivity of TM and TE modes is larger than in isotropic materials which will help to implement schemes based on relative frequency measurements. Finally \mgf resonators are known for their high $Q$ factors \cite{alnis_thermal-noise_2011, liang_generation_2011}, which do not degraded in aqueous environment \cite{tavernier_magnesium_2010}, apart from the influence of potential absorption of the evanescent field by water.

In the first part of the paper we discuss the differences concerning sensitivity, behavior of TM and TE modes and evanescent field length between fused silica and \mgf resonators theoretically. In the second part we present experimental results showing the significantly extended evanescent field decay length and sensitivities against refractive index changes of two different \mgf WGM resonator geometries.

\section{Theory}

Whispering gallery modes have been analytically described for different geometries in various publications \cite{schiller_high-resolution_1991, oraevsky_whispering-gallery_2002, gorodetsky_geometrical_2006, demchenko_analytical_2013, breunig_whispering_2013}. Such modes arise from boundary conditions at the rim of convex shaped, usually rotationally symmetric, dielectric bodies like spheres or spheroids. Due to the resonator's higher refractive index compared with the surrounding, the light can be guided and confined via total internal reflection at the boundary. Geometry and material properties determine the resonance positions of the different types of modes. Usually these modes can be characterized by the three mode numbers $l, q, p$ which correspond to different intensity distributions of the light close to the boundary: $q \in \mathbb{N}$ counts the number of maxima in radial direction, $p + 1 \in \mathbb{N}$ corresponds to the number of maxima in polar direction and $l-p$ is the number of field oscillations along one roundtrip around the cavity. For close to fundamental modes $l$ can be estimated sufficiently with $l \approx 2 \pi R n_r/\lambda_0$.

The spectral position of the modes are found by solving Maxwell's equations under proper boundary conditions. They depend on the mode numbers, the refractive index of the resonator $n_r$ and its radius $R$ as well as the index of the surrounding $n_s$. For refractometric sensing, we are interested in the dependence of the resonance wavelengths on small changes of $n_s$, which can be easily calculated by taking the derivative of the modal dispersion relation (see Ref. \cite{hanumegowda_refractometric_2005}). This yields for a spherical geometry
\begin{equation}
\frac{\partial}{\partial n_s} \lambda^{(TE)}_{l,q} = \frac{\lambda^2}{2 \pi R} \frac{n_s}{\left(n_r^2 - n_s^2\right)^{3/2}} \left[1 - \frac{\zeta_q}{2^{1/3}} \frac{n_r^2}{n_r^2 - n_s^2} \left(l+\frac{1}{2}\right)^{-2/3}\right],
\label{equ:TE}
\end{equation}

\begin{equation}
\begin{split}
\frac{\partial}{\partial n_s} \lambda^{(TM)}_{l,q} = & \frac{\lambda^2}{2 \pi R} \frac{n_s}{n_r^2 \left(n_r^2 - n_s^2\right)^{3/2}} \left[\vphantom{\frac{a^2}{b^2}} 2 n_r^2 - n_s^2  \right.\\
& \left. - \frac{\zeta_q}{2^{1/3}} \frac{2 n_r^6 + n_r^4 n_s^2 - 4 n_r^2 n_s^4 + 2 n_s^6}{n_r^2 \left(n_r^2 - n_s^2 \right)} \left(l+\frac{1}{2} \right)^{-2/3}\right]
\end{split}
\label{equ:TM}
\end{equation}
The superscript denotes the polarization, where TE corresponds to modes with electric field vectors parallel to the surface and TM perpendicular to it. Due to different boundary conditions the two polarizations have slightly different sensitivities, even for isotropic resonator materials such as fused silica.
Note that there is no dependence on the mode number $p$. This degeneracy is a consequence of the radial symmetry of a perfect sphere and is usually not observed in experiments due to imperfections in the fabrication process of the resonator. If the deviation from a sphere is strong, which is commonly the case for mechanically produced crystalline WGM resonators, the $p \neq 0$ modes have to be described by a slightly modified dispersion relation which can be found in \cite{gorodetsky_geometrical_2006, demchenko_analytical_2013, breunig_whispering_2013}. These equations contain terms depending on the eccentricity which, however, do not depend on the index of refraction and therefore cancel out when taking the derivative. Thus we conclude that Eq. (\ref{equ:TE}) and (\ref{equ:TM}) predict the sensitivity of non spherical resonators satisfactorily.

\begin{figure}[t]
\centering\includegraphics[height=7cm]{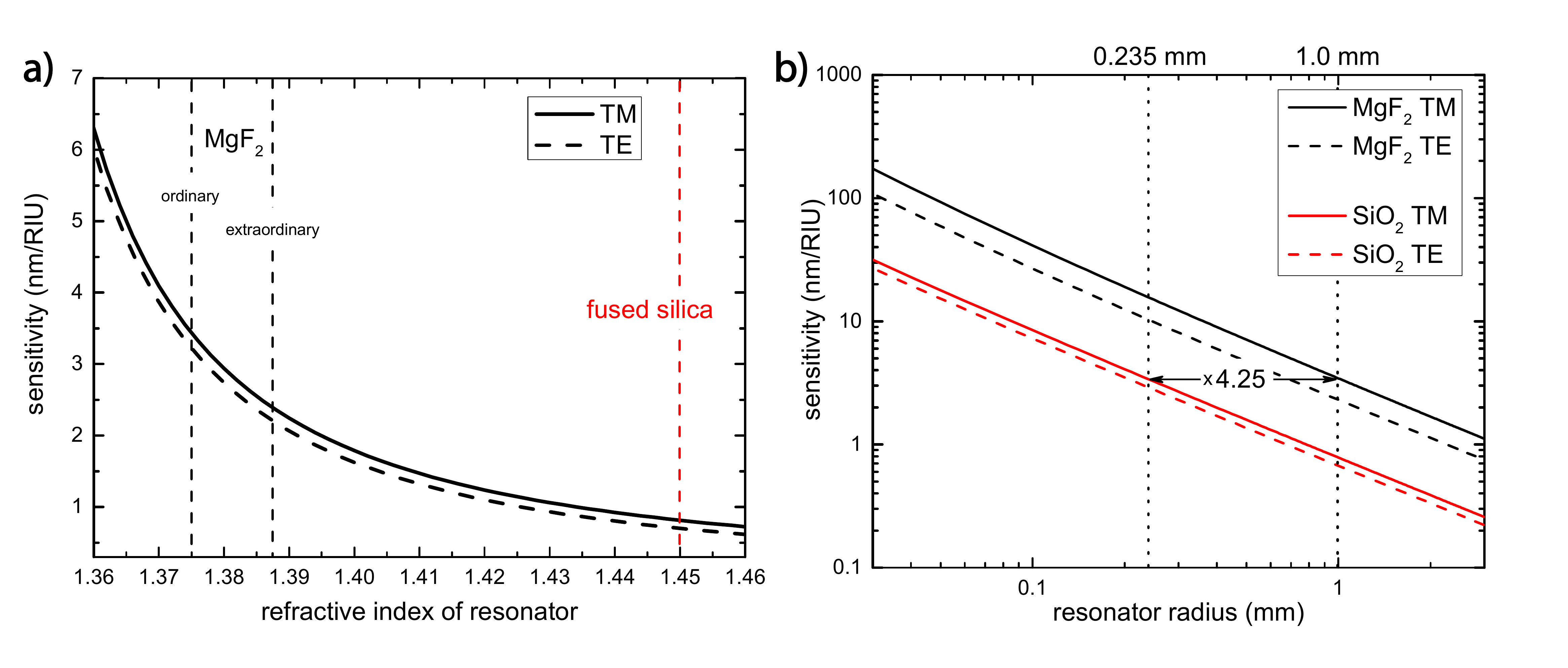} 
\caption{a) shows the dependence of the sensitivity (wavelength change of the modes over refractive index unit (RIU) change) on the bulk index of the used resonator material. The resonator radius was assumed to be $R = 1$\,mm and we only consider fundamental modes ($q = 1$). TM and TE modes have slightly different sensitivities due to different boundary conditions, but both increase when the resonator index comes closer to the index of the surrounding (which is assumed to be water, $n_s = 1.329$ at $795$\,nm). The refractive indices of \mgf and fused silica are marked by dashed lines. It is apparent that the birefringence of \mgf causes a stronger dispersion between TE and TM as compared to the isotropic cases such as fused silica. b) shows the sensitivities of \mgf and fused silica resonators against their radius. The linearity in the double logarithmic plot indicates that \mgf resonators can always be $4.25$ times larger to reach the same sensitivity as fused silica ones. Moreover the ratio between TM and TE sensitivities stays constant (nameley $1.49$ for \mgf and $1.17$ for fused silica).}
\label{fig:sensitivity_theory}
\end{figure}

For the following considerations we always assume fundamental modes with $q = 1$ and \mbox{$p=1$}, however, it is worth mentioning that the sensitivity increases by approximately $2-3\%$ per increased $q$-number. Apart from that, the sensitivity is mainly determined by the refractive index difference between the resonator and the surrounding $n_r^2 - n_s^2$ and the radius $R$. This is illustrated in Fig.~\ref{fig:sensitivity_theory} where Eq.~(\ref{equ:TE}) and (\ref{equ:TM}) are plotted for different geometries and materials. In Fig.~\ref{fig:sensitivity_theory}(a) the index of the surrounding is assumed as $n_s = 1.329$ (water at $795$\,nm) and the sensitivity is plotted against the index of the resonator $n_r$. It is apparent that the sensitivity increases significantly when $n_r$ approaches $n_s$. The plot also shows that the performance of \mgf is in general better than fused silica due to its smaller refractive index. In addition \mgf is slightly birefringent ($n_o = 1.375$, $n_e = 1.387$, see \cite{dodge_refractive_1984}), hence the difference between TE and TM modes (corresponding to extraordinary and ordinary polarization in a $z$-cut resonator) is stronger than in isotropic fused silica. In Fig.~\ref{fig:sensitivity_theory}(b) we plot the sensitivity against the resonator radius for \mgf and fused silica for both polarizations. The double logarithmic plot reveals that the sensitivity scales nearly inversely proportional for all cases down to very small resonator sizes. Therefore we find, that a \mgf resonator can be $4.25$ times larger than a fused silica one to show the same sensitivity against refractive index change. It is interesting to mention, that the ratio of shifts between TM and TE modes is also nearly constant for all cases. They are $1.17$ for fused silica and $1.49$ for \mgf WGM resonators.

The higher sensitivity of \mgf has certain advantages over fused silica. As the resonator can be approximately four times larger to provide the same performance, a mechanically stable implementation is much easier. Also, larger resonators show less noise due to thermally induced short term fluctuations. Significant contributions come either from thermoelasticity or thermorefraction which scale inversely proportional to the resonator volume and the modal volume respectively \cite{gorodetsky_fundamental_2004, savchenkov_whispering-gallery-mode_2007}. However, we want to point out, that long term thermal drifts, which are the major problem in most resonator based sensing applications are independent of the resonator size and arise only from material parameters such as thermal expansion and thermorefraction. These are very similar to those of fused silica and thus thermal drifts are expected to be in the same order of magnitude. To overcome this problem, Le et al. \cite{le_optical_2009} proposed and demonstrated a scheme using difference frequency measurements between TE and TM modes in fused silica. As this difference is significantly more sensitive to refractive index changes than to temperature changes, the influence of thermal drifts has been reduced. Such a scheme requires high enough modal density, to observe TE and TM modes within the sweeping range of the laser simultaneously, and the difference between the TE and TM sensitivities being as large as possible. Both is given in \mgf resonators due to the larger resonator size and the materials birefringence.

Finally, we want to point out that the small index contrast of \mgf with respect to water results in a significantly longer decay distance of the evanescent field leaking out of the cavity. The field decay outside the dielectric $r > R$ can be described as \cite{little_analytic_1999}
\begin{equation}
E(r) \propto e^{-\kappa r} \quad \text{with} \quad \kappa \approx \frac{2 \pi }{\lambda_0} \sqrt{n_{\mathrm{eff}}^2 - n^2_s}
\end{equation}
here, $\lambda_0$ is the vacuum wavelength and $n_{\mathrm{eff}}$ is the effective refractive index of the considered mode, which is always below, but usually close to the bulk refractive index $n_r$. From this the field decay length in water $\kappa^{-1}$ can be calculated to be $379$\,nm for TM in a \mgf resonator with $R = 1$\,mm and only $231$\,nm for TM in a fused silica sphere with $R = 0.235$\,mm. The radii correspond to resonators of same bulk index sensitivity. Such a long evanescent field extension can be advantageous because the overlap and thus the sensitivity will be enhanced if sensing of large particles or microorganisms like bacteria is desired \cite{ren_high-q_2007}.

\section{Experiment}

\begin{figure}[t]
\centering\includegraphics[height=6cm]{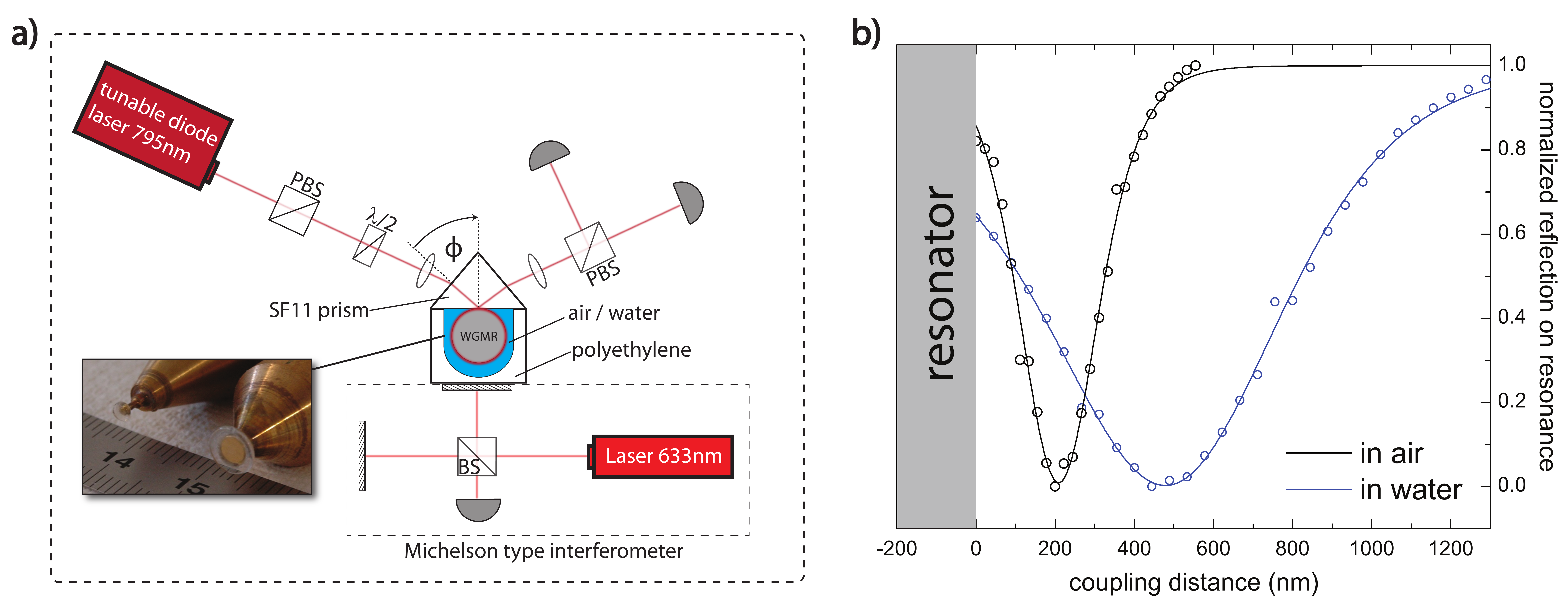}
\caption{a) shows the setup used to characterize the two \mgf resonators. They can be mounted from above within a basin which itself is mounted onto a piezostage. One of the walls of the basin is a SF11 prism which is used for coupling a tunable laser around $795$\,nm to TM and TE modes simultaneously by using $45^\circ$ polarized light. The reflected light is separated via a polarizing beam splitter and sent to two photodiodes to enable tracing of TM and TE spectra at the same time. A mirror which is part of a Michelson type interferometer is attached to the backside of the basin to measure the displacement of the prism with respect to the resonator. The results of this measurement are shown in b): we plot the normalized coupling contrast of two (different) TM modes of the large resonator ($R = 2.91$\,mm) in air and water against the traveling distance of the coupling prism. From this we extract the evanescent field decay length for air (water) $\kappa^{-1} = 134 (382)$\,nm by fitting equation (\ref{equ:coupling}) to the data.}
\label{fig:Setup_Coupling}
\end{figure}

The experimental setup is depicted in Fig.~\ref{fig:Setup_Coupling}(a). The resonator is glued onto a rigid brass rod which is mounted `head down' such that the resonator is located in a basin which can be water filled. Three of the basin's walls are made of polyethylene while the fourth is a SF11 glass prism ($n_p \approx 1.77$) which is used for evanescent coupling to the resonator. The light, coming from a widely tunable grating stabilized diode laser emitting around $795$\,nm, was adjusted to $45^\circ$ linear polarization to excite TE and TM modes simultaneously. In reflection, the light is separated by a polarizing beamsplitter to discriminate and trace TE and TM modes at the same time.
The coupling strength can be adjusted by moving the whole basin and thus the prism via a piezo stage. To investigate the evanescent field decay, we measured the traveling distance of the piezo via a Michelson type interferometer whereby one of the mirrors is attached to the backside of the basin.

For our experiments, we use two \mgf resonators ($n_o = 1.375$, $n_e = 1.387$ at $795$\,nm) of different size. Both were fabricated on a home build lathe as described in \cite{grudinin_ultrahigh_2006}. A diamond cutter was used to preshape the rim of the resonators for optimized incoupling followed by careful hand polishing with a diamond slurry to ensure optical quality of the surface and hence high $Q$ factors. The larger resonator has a radius of $R=2.91$\,mm and the smaller one $R=1.19$\,mm.

Due to the birefringence of \mgf, the fundamental TE and TM modes have slightly different coupling angles according to $\phi = \arcsin(n_r/n_p)$ (see Fig.~\ref{fig:Setup_Coupling}(a)) inside the prism. The optimal coupling angles outside the prism hence differ by about $1.1^\circ$ which is small enough to excite fundamental modes of both polarizations simultaneously with minor loss of coupling efficiency.

\begin{figure}[t]
\centering\includegraphics[height=8cm]{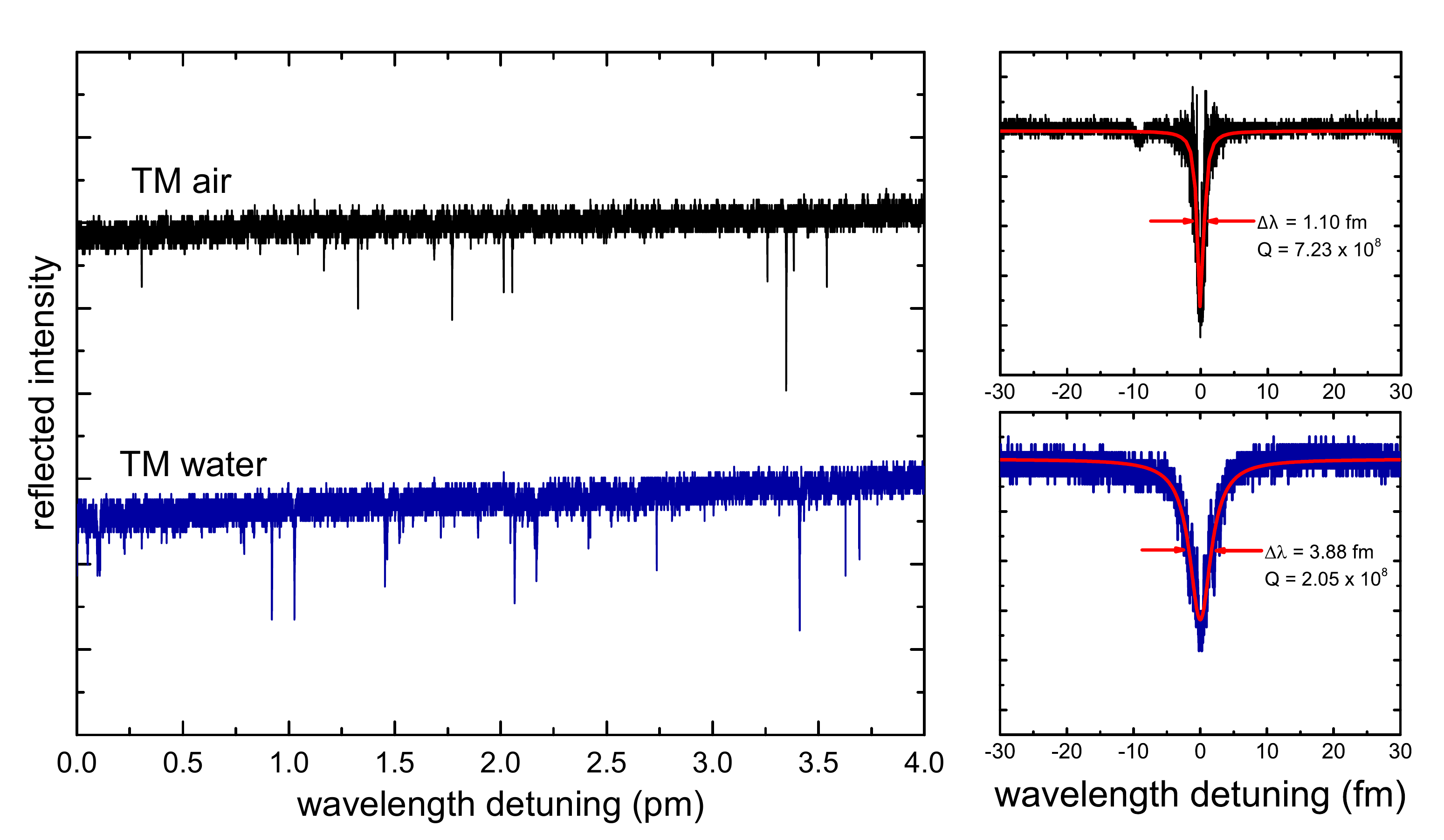}
\caption{shows exemplary overview spectra of different TM modes of the large resonator in air and immersed in water after readjusting coupling. Qualitatively, modal density and coupling efficiency remains the same in water. Also shown are zoom ins of (two different) TM modes before and after immersing the resonator. The loaded $Q$ factor was derived from the linewidth acquired by Lorentzian fits while the resonator was critically coupled. The measured linewidth in air was limited by the linewidth of the used laser system.}
\label{fig:Spectra}
\end{figure}

Figure~\ref{fig:Spectra} shows exemplary overview spectra and $Q$ factor measurements for TM modes of the large resonator in air and in deionized water. Loaded quality factors were derived from sideband calibrated line width measurements at critical coupling according to $Q = \nu / \Delta \nu$. 
While coupling depth and modal density remain approximately the same when the resonators are immersed in water, the $Q$ factor drops significantly as water has higher absorption than \mgf at 795\,nm. For TM (TE) we measured $Q_\mathrm{air} = 7.3 \times 10^8$ ($1.9 \times 10^8$) and $Q_\mathrm{water} = 2.1 \times 10^8$ ($1.7 \times 10^8$). The $Q$ factors for the small resonator are slightly smaller, but still remain above $1 \times 10^8$ in air and water. Apart from the initial drop, the $Q$ factors were stable over several hours in water.

We investigated the length of the evanescent field by measuring the intensity reflected from the cavity at resonance wavelength for different coupling strengths while also registering the traveling distance of the prism via the interferometer. In Fig.~\ref{fig:Setup_Coupling}(b) such data for the small resonator in air and water environment are plotted for two (different) TM modes: the reflected intensity is normalized such that critical coupling (maximum contrast) corresponds to zero and no coupling to one. The normalized reflected intensity can be described by \cite{vyatchanin_tunable_1992}

\begin{equation}
\mathfrak{R} = \left( \frac{1-A r e^{-2 \kappa d}}{1 + r e^{-2 \kappa d}} \right)^2,
\label{equ:coupling}
\end{equation}
where $A$ contains the geometrical matching of the incoming beam with the outcoupled one, while $r = \delta_c / \delta_i$ is the ratio between maximally achievable coupling $\delta_c$ and intrinsic loss $\delta_i$ and $d$ is the distance of the coupler to the resonator's surface. From this fit we derive the field decay lengths $\kappa^{-1} = 134$\,nm in air and $382$\,nm in water which matches the theoretical values of $135$\,nm and $377$\,nm quite well. Note that the theoretical values were derived using $n_{\mathrm{eff}}$ for the a fundamental mode $q=1$ and not just the bulk index.
\begin{figure}[t]
\centering\includegraphics[height=11cm]{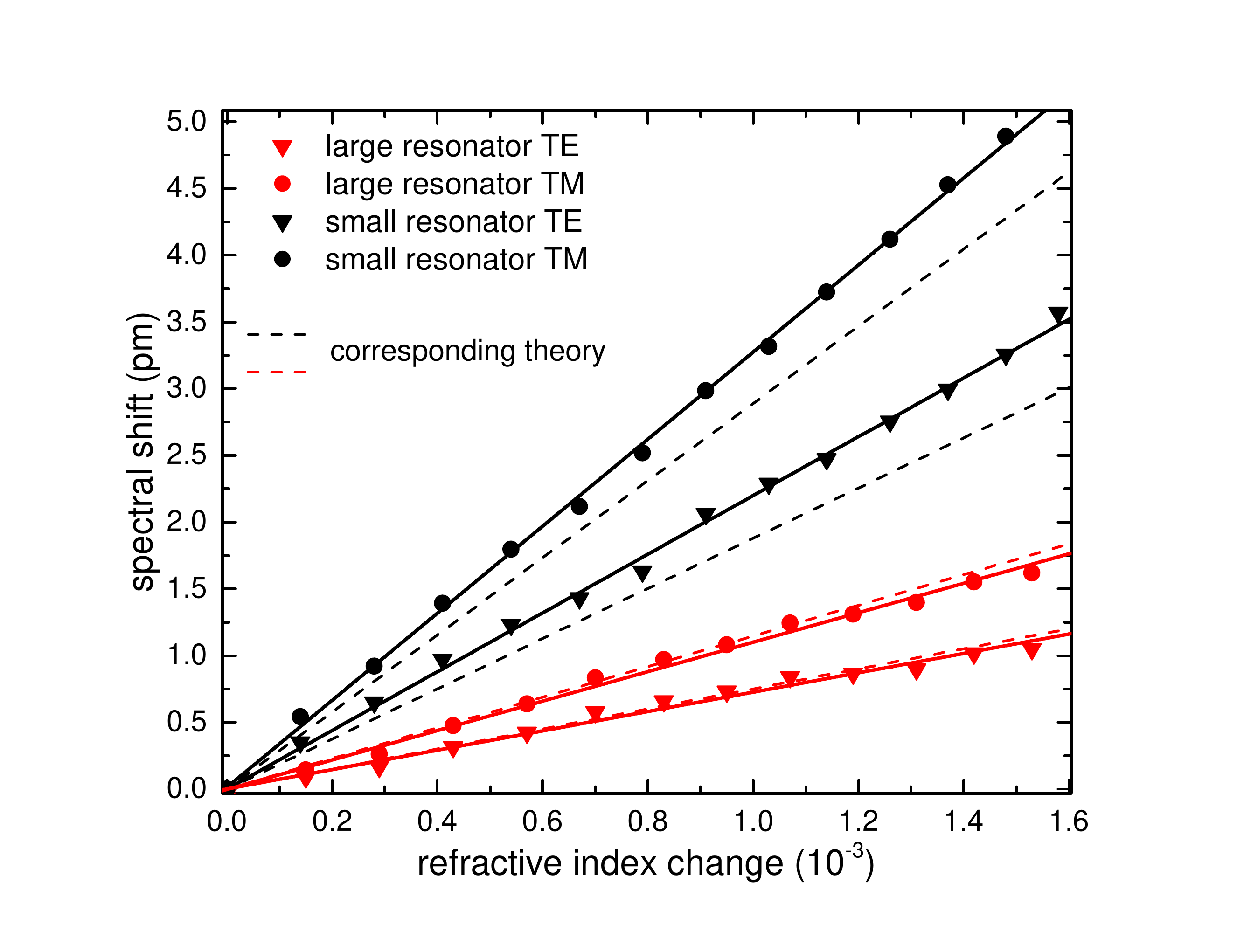}
\caption{The plot shows the measured wavelength change of TE and TM modes against refractive index change of the surrounding water for the large (R = 2.91\,mm) and the small (R = 1.19\,mm) resonator. Linear fits, indicated by the straight lines, where used to extract the sensitivities. The dashed lines correspond to the expected responses derived from Eqs.~(\ref{equ:TE}) and (\ref{equ:TM}) for the different cases. One can see that the sensitivity of the small resonator is about $15\%$ above the theoretically predicted trend.}
\label{fig:Sensitivity}
\end{figure}

For sensitivity measurements, we filled the basin with a defined amount of deionized water, waited till the thermal equilibrium was reached and readjusted the resonator-prism distance to reach the critical coupling regime again. We prepared a glycerol solution of known concentration and also let it reach equilibrium conditions. For each measurement, \SI{10}{\micro \litre} of the glycerol solution were added to the basin and mixed carefully to prevent the buildup of concentration gradients. After 60\,s, when the mode shifts reached equilibrium, both spectra, TE and TM, where acquired via an oscilloscope. The shifts where evaluated by fitting a Lorenzian to each mode and comparing the peak positions between each measurement. We calculate the refractive index change for each measurement from the injected amount of glycerol using the Lorentz-Lorenz relation \cite{herraez_refractive_2006}. The results are presented in Fig.~\ref{fig:Sensitivity} where the wavelength shift of modes of both polarizations from both resonators are plotted against the refractive index change. As this change is only small, the spectrum's response is as expected for both resonators linear and the sensitivity can be extracted from the slope of a linear fit to each curve.

Qualitatively the curves match the expected behavior: the small resonator is more sensitive to refractive index changes than the larger one and in both resonators the TM mode shifts stronger than the TE mode. The large resonator shows a sensitivity of $1.10\,(0.73)$\,nm/RIU for TM (TE) which deviates less than $5\%$ from the theoretical values for $q = 1$ modes. The small resonator, however, shows a sensitivity of $3.26\, (2.19)$\,nm/RIU for TM (TE), which is about $15\%$ higher than theory predicts for fundamental modes. In Fig.~\ref{fig:Sensitivity} the theoretically expected responses for the small resonator are represented by the dashed lines and show that the experimental sensitivity exceeds the theoretical one systematically. 

\section{Conclusion}

In this paper, we discuss and demonstrate for the first time the feasibility of high-$Q$ crystalline whispering gallery mode resonators made from magnesium fluoride for refractometric sensing. We show theoretically that \mgf with its refractive index being close to that of water and slightly polarization dependent (birefringent) has some advantageous properties over the standard sensing material fused silica. The smaller refractive index leads to stronger interaction with the surrounding medium and thus to better sensitivity, which allows to use larger resonators providing intrinsically larger modal density and better short term stability than a fused silica resonator of same sensitivity. These properties simplify rigid mechanical mounting, the implementation of measurement schemes requiring multiple modes and the tracing of fast processes. The birefringence amplifies the sensitivity difference between TM and TE mode families which is helpful for the implementation of temperature drift neglecting schemes.

We do first experiments, proving that \mgf WGM resonators are suitable devices for refractometric applications. Two resonators of different size in the mm regime were fabricated via diamond turning and carefully characterized in air and aqueous environment. In water, both resonators show non decaying loaded $Q$-factors above $10^8$ and keep their high modal density. Using interferometric distance measurements, we show that the penetration depth of the evanescent field matches the theoretical prediction quite well and is thus approximately two times larger than for fused silica resonators of same sensitivity. Finally, the sensitivity dependence on polarization and resonator size is demonstrated by measuring the response of TM and TE modes to small refractive index changes of the water. Here we observe a higher than expected sensitivity for the small resonator. The reason for this is unclear and will be subject of a further study.

Theoretical calculations \cite{little_analytic_1999} show that even for the small index difference between \mgf and water the radiation limited $Q$ factor remains far above $10^8$ for resonator radii smaller than $250$\,\si{\micro\m}. This means, according to our calculations and measurements, that \mgf resonators can easily compete with fused silica spheres smaller than $60$\,\si{\micro\m}, while maintaining the advantages we have pointed out.
Furthermore, we recently showed that the slight birefringence of a \mgf WGM resonator can harbor elliptically polarized modes if its optic axis is tilted with respect to the rotational symmetry axis while preserving ultrahigh $Q$ factors \cite{sedlmeir_experimental_2013}. This feature could enable sensing of optically active liquids.

\section{Acknowledgements}
We acknowledge fruitful discussions with Martin D.  Baaske, Matthew R. Foreman and Frank Vollmer from the Lab of Nanophotonics and Biosensing. We also acknowledge support by Deutsche Forschungsgemeinschaft and Friedrich-Alexander-Universit\"at Erlangen-N\"urnberg (FAU) within the funding programme Open Access Publishing.
\bibliography{Sensing3}

\begin{thebibliography}{29}
\providecommand{\natexlab}[1]{#1}
\providecommand{\url}[1]{\texttt{#1}}
\expandafter\ifx\csname urlstyle\endcsname\relax
  \providecommand{\doi}[1]{doi: #1}\else
  \providecommand{\doi}{doi: \begingroup \urlstyle{rm}\Url}\fi

\bibitem[Alnis et~al.(2011)Alnis, Schliesser, Wang, Hofer, Kippenberg, and
  Hänsch]{alnis_thermal-noise_2011}
J.~Alnis, A.~Schliesser, C.~Y Wang, J.~Hofer, T.~J Kippenberg, and T.~W
  Hänsch.
\newblock Thermal-noise limited laser stabilization to a crystalline
  whispering-gallery-mode resonator.
\newblock \emph{1102.4227}, February 2011.
\newblock URL \url{http://arxiv.org/abs/1102.4227}.

\bibitem[Arnold et~al.(2003)Arnold, Khoshsima, Teraoka, Holler, and
  Vollmer]{arnold_shift_2003}
S.~Arnold, M.~Khoshsima, I.~Teraoka, S.~Holler, and F.~Vollmer.
\newblock Shift of whispering-gallery modes in microspheres by protein
  adsorption.
\newblock \emph{Opt. Lett.}, 28\penalty0 (4):\penalty0 272--274, February 2003.
\newblock \doi{10.1364/OL.28.000272}.
\newblock URL \url{http://ol.osa.org/abstract.cfm?URI=ol-28-4-272}.

\bibitem[Breunig et~al.(2013)Breunig, Sturman, Sedlmeir, Schwefel, and
  Buse]{breunig_whispering_2013}
I.~Breunig, B.~Sturman, F.~Sedlmeir, H.~G.~L. Schwefel, and K.~Buse.
\newblock Whispering gallery modes at the rim of an axisymmetric optical
  resonator: Analytical versus numerical description and comparison with
  experiment.
\newblock \emph{Opt. Express}, 21\penalty0 (25):\penalty0 30683--30692,
  December 2013.
\newblock \doi{10.1364/OE.21.030683}.
\newblock URL
  \url{http://www.opticsexpress.org/abstract.cfm?URI=oe-21-25-30683}.

\bibitem[Dantham et~al.(2013)Dantham, Holler, Barbre, Keng, Kolchenko, and
  Arnold]{dantham_label-free_2013}
Venkata~R. Dantham, Stephen Holler, Curtis Barbre, David Keng, Vasily
  Kolchenko, and Stephen Arnold.
\newblock Label-free detection of single protein using a nanoplasmonic-photonic
  hybrid microcavity.
\newblock \emph{Nano Lett.}, 13\penalty0 (7):\penalty0 3347--3351, July 2013.
\newblock ISSN 1530-6984.
\newblock \doi{10.1021/nl401633y}.
\newblock URL \url{http://dx.doi.org/10.1021/nl401633y}.

\bibitem[Demchenko and Gorodetsky(2013)]{demchenko_analytical_2013}
Yury~A. Demchenko and Michael~L. Gorodetsky.
\newblock Analytical estimates of eigenfrequencies, dispersion, and field
  distribution in whispering gallery resonators.
\newblock \emph{J. Opt. Soc. Am. B}, 30\penalty0 (11):\penalty0 3056--3063,
  November 2013.
\newblock \doi{10.1364/JOSAB.30.003056}.
\newblock URL \url{http://josab.osa.org/abstract.cfm?URI=josab-30-11-3056}.

\bibitem[Dodge(1984)]{dodge_refractive_1984}
Marilyn~J. Dodge.
\newblock Refractive properties of magnesium fluoride.
\newblock \emph{Appl. Opt.}, 23\penalty0 (12):\penalty0 1980, June 1984.
\newblock ISSN 0003-6935, 1539-4522.
\newblock \doi{10.1364/AO.23.001980}.
\newblock URL
  \url{http://www.opticsinfobase.org/ao/fulltext.cfm?uri=ao-23-12-1980&id=27584}.

\bibitem[Gorodetsky and Grudinin(2004)]{gorodetsky_fundamental_2004}
Michael~L. Gorodetsky and Ivan~S. Grudinin.
\newblock Fundamental thermal fluctuations in microspheres.
\newblock \emph{J. Opt. Soc. Am. B}, 21\penalty0 (4):\penalty0 697–705, April
  2004.
\newblock \doi{10.1364/JOSAB.21.000697}.
\newblock URL \url{http://josab.osa.org/abstract.cfm?URI=josab-21-4-697}.

\bibitem[Gorodetsky and Fomin(2006)]{gorodetsky_geometrical_2006}
M.L. Gorodetsky and A.E. Fomin.
\newblock Geometrical theory of whispering-gallery modes.
\newblock \emph{IEEE J. Sel. Top. Quant. Electron.}, 12\penalty0 (1):\penalty0
  33--39, 2006.
\newblock ISSN 1077-260X.
\newblock \doi{10.1109/JSTQE.2005.862954}.
\newblock URL \url{10.1109/JSTQE.2005.862954}.

\bibitem[Grudinin et~al.(2006)Grudinin, Ilchenko, and
  Maleki]{grudinin_ultrahigh_2006}
Ivan~S. Grudinin, Vladimir~S. Ilchenko, and Lute Maleki.
\newblock {Ultrahigh optical Q factors of crystalline resonators in the linear
  regime}.
\newblock \emph{Phys. Rev. A}, 74\penalty0 (6):\penalty0 063806, December 2006.
\newblock \doi{10.1103/PhysRevA.74.063806}.
\newblock URL \url{http://link.aps.org/doi/10.1103/PhysRevA.74.063806}.

\bibitem[Guan et~al.(2006)Guan, Arnold, and Otugen]{guan_temperature_2006}
Guoming Guan, Stephen Arnold, and Volkan Otugen.
\newblock Temperature measurements using a microoptical sensor based on
  whispering gallery modes.
\newblock \emph{{AIAA} Journal}, 44\penalty0 (10):\penalty0 2385--2389, 2006.
\newblock ISSN 0001-1452.
\newblock \doi{10.2514/1.20910}.
\newblock URL \url{http://arc.aiaa.org/doi/abs/10.2514/1.20910}.

\bibitem[Hanumegowda et~al.(2005)Hanumegowda, Stica, Patel, White, and
  Fan]{hanumegowda_refractometric_2005}
Niranjan~M. Hanumegowda, Caleb~J. Stica, Bijal~C. Patel, Ian White, and Xudong
  Fan.
\newblock Refractometric sensors based on microsphere resonators.
\newblock \emph{Appl. Phys. Lett.}, 87\penalty0 (20):\penalty0
  201107–201107--3, 2005.
\newblock ISSN 0003-6951.
\newblock \doi{10.1063/1.2132076}.

\bibitem[Herráez and Belda(2006)]{herraez_refractive_2006}
Jose~V. Herráez and R.~Belda.
\newblock Refractive indices, densities and excess molar volumes of
  monoalcohols + water.
\newblock \emph{J. Solution Chem.}, 35\penalty0 (9):\penalty0 1315--1328,
  September 2006.
\newblock ISSN 0095-9782, 1572-8927.
\newblock \doi{10.1007/s10953-006-9059-4}.
\newblock URL \url{http://link.springer.com/article/10.1007/s10953-006-9059-4}.

\bibitem[Hunt et~al.(2010)Hunt, Soteropulos, and
  Armani]{hunt_bioconjugation_2010}
Heather~K. Hunt, Carol Soteropulos, and Andrea~M. Armani.
\newblock Bioconjugation strategies for microtoroidal optical resonators.
\newblock \emph{Sensors}, 10\penalty0 (10):\penalty0 9317--9336, October 2010.
\newblock \doi{10.3390/s101009317}.
\newblock URL \url{http://www.mdpi.com/1424-8220/10/10/9317}.

\bibitem[Ioppolo et~al.(2008)Ioppolo, Kozhevnikov, Stepaniuk, \"Ot\"ugen, and
  Sheverev]{ioppolo_micro-optical_2008}
Tindaro Ioppolo, Michael Kozhevnikov, Vadim Stepaniuk, M.~Volkan \"Ot\"ugen,
  and Valery Sheverev.
\newblock Micro-optical force sensor concept based on whispering gallery mode
  resonators.
\newblock \emph{Appl. Opt.}, 47\penalty0 (16):\penalty0 3009--3014, June 2008.
\newblock \doi{10.1364/AO.47.003009}.
\newblock URL \url{http://ao.osa.org/abstract.cfm?URI=ao-47-16-3009}.

\bibitem[Le et~al.(2009)Le, Savchenkov, Yu, Maleki, and
  Steier]{le_optical_2009}
Thanh Le, Anatoliy Savchenkov, Nan Yu, Lute Maleki, and W.~H. Steier.
\newblock Optical resonant sensors: a method to reduce the effect of thermal
  drift.
\newblock \emph{Appl. Opt.}, 48\penalty0 (3):\penalty0 458–463, January 2009.
\newblock \doi{10.1364/AO.48.000458}.
\newblock URL \url{http://ao.osa.org/abstract.cfm?URI=ao-48-3-458}.

\bibitem[Liang et~al.(2011)Liang, Savchenkov, Matsko, Ilchenko, Seidel, and
  Maleki]{liang_generation_2011}
W.~Liang, A.~A. Savchenkov, A.~B. Matsko, V.~S. Ilchenko, D.~Seidel, and
  L.~Maleki.
\newblock Generation of near-infrared frequency combs from a {MgF}2 whispering
  gallery mode resonator.
\newblock \emph{Optics Letters}, 36\penalty0 (12):\penalty0 2290--2292, June
  2011.
\newblock \doi{10.1364/OL.36.002290}.
\newblock URL \url{http://ol.osa.org/abstract.cfm?URI=ol-36-12-2290}.

\bibitem[Little et~al.(1999)Little, Laine, and Haus]{little_analytic_1999}
{B.E.} Little, J.-P. Laine, and {H.A.} Haus.
\newblock Analytic theory of coupling from tapered fibers and half-blocks into
  microsphere resonators.
\newblock \emph{J. Lightwave Technol.}, 17\penalty0 (4):\penalty0 704–715,
  1999.
\newblock ISSN 0733-8724.
\newblock \doi{10.1109/50.754802}.
\newblock URL \url{10.1109/50.754802}.

\bibitem[Lutti et~al.(2008)Lutti, Langbein, and Borri]{lutti_monolithic_2008}
Julie Lutti, Wolfgang Langbein, and Paola Borri.
\newblock A monolithic optical sensor based on whispering-gallery modes in
  polystyrene microspheres.
\newblock \emph{Appl. Phys. Lett.}, 93\penalty0 (15):\penalty0 151103, 2008.
\newblock ISSN 00036951.
\newblock \doi{10.1063/1.2998652}.
\newblock URL \url{http://orca.cf.ac.uk/9041/}.

\bibitem[Oraevsky(2002)]{oraevsky_whispering-gallery_2002}
Anatolii~N Oraevsky.
\newblock Whispering-gallery waves.
\newblock \emph{Quant. Electron}, 32\penalty0 (5):\penalty0 377--400, May 2002.
\newblock ISSN 1063-7818.
\newblock \doi{10.1070/QE2002v032n05ABEH002205}.
\newblock URL
  \url{http://www.turpion.org/php/paper.phtml?journal_id=qe&paper_id=2205}.

\bibitem[Ren et~al.(2007)Ren, Vollmer, Arnold, and Libchaber]{ren_high-q_2007}
Hai-Cang Ren, Frank Vollmer, Stephen Arnold, and Albert Libchaber.
\newblock {High-Q microsphere biosensor - analysis for adsorption of rodlike
  bacteria}.
\newblock \emph{Opt. Express}, 15\penalty0 (25):\penalty0 17410--17423,
  December 2007.
\newblock ISSN 1094-4087.
\newblock {PMID:} 19551035.

\bibitem[Savchenkov et~al.(2007)Savchenkov, Matsko, Ilchenko, Yu, and
  Maleki]{savchenkov_whispering-gallery-mode_2007}
Anatoliy~A. Savchenkov, Andrey~B. Matsko, Vladimir~S. Ilchenko, Nan Yu, and
  Lute Maleki.
\newblock Whispering-gallery-mode resonators as frequency references. {II.}
  stabilization.
\newblock \emph{J. Opt. Soc. Am. B}, 24\penalty0 (12):\penalty0 2988–2997,
  December 2007.
\newblock \doi{10.1364/JOSAB.24.002988}.
\newblock URL \url{http://josab.osa.org/abstract.cfm?URI=josab-24-12-2988}.

\bibitem[Schiller and Byer(1991)]{schiller_high-resolution_1991}
Stephan Schiller and R.~L. Byer.
\newblock High-resolution spectroscopy of whispering gallery modes in large
  dielectric spheres.
\newblock \emph{Opt. Lett.}, 16\penalty0 (15):\penalty0 1138–1140, August
  1991.
\newblock \doi{10.1364/OL.16.001138}.
\newblock URL \url{http://ol.osa.org/abstract.cfm?URI=ol-16-15-1138}.

\bibitem[Sedlmeir et~al.(2013)Sedlmeir, Hauer, Fürst, Leuchs, and
  Schwefel]{sedlmeir_experimental_2013}
Florian Sedlmeir, Martin Hauer, Josef~U. Fürst, Gerd Leuchs, and Harald G.~L.
  Schwefel.
\newblock Experimental characterization of an uniaxial angle cut whispering
  gallery mode resonator.
\newblock \emph{Optics Express}, 21\penalty0 (20):\penalty0 23942--23949,
  October 2013.
\newblock \doi{10.1364/OE.21.023942}.
\newblock URL
  \url{http://www.opticsexpress.org/abstract.cfm?URI=oe-21-20-23942}.

\bibitem[Tavernier et~al.(2010)Tavernier, Salzenstein, Volyanskiy, Chembo, and
  Larger]{tavernier_magnesium_2010}
H.~Tavernier, P.~Salzenstein, K.~Volyanskiy, Y.K. Chembo, and L.~Larger.
\newblock Magnesium fluoride whispering gallery mode disk-resonators for
  microwave photonics applications.
\newblock \emph{{IEEE} Photonics Technology Letters}, 22\penalty0
  (22):\penalty0 1629--1631, November 2010.
\newblock ISSN 1041-1135.
\newblock \doi{10.1109/LPT.2010.2075923}.

\bibitem[Vollmer et~al.(2008)Vollmer, Arnold, and Keng]{vollmer_single_2008}
F.~Vollmer, S.~Arnold, and D.~Keng.
\newblock Single virus detection from the reactive shift of a
  whispering-gallery mode.
\newblock \emph{Proc. Natl. Acad. Sci. U.S.A.}, 105\penalty0 (52):\penalty0
  20701--20704, December 2008.
\newblock ISSN 0027-8424, 1091-6490.
\newblock \doi{10.1073/pnas.0808988106}.
\newblock URL \url{http://www.pnas.org/content/105/52/20701}.
\newblock {PMID:} 19075225.

\bibitem[Vollmer and Arnold(2008)]{vollmer_whispering-gallery-mode_2008}
Frank Vollmer and Stephen Arnold.
\newblock Whispering-gallery-mode biosensing: label-free detection down to
  single molecules.
\newblock \emph{Nat. Methods}, 5\penalty0 (7):\penalty0 591--596, July 2008.
\newblock ISSN 1548-7091.
\newblock \doi{10.1038/nmeth.1221}.
\newblock URL
  \url{http://www.nature.com/nmeth/journal/v5/n7/abs/nmeth.1221.html}.

\bibitem[Vyatchanin et~al.(1992)Vyatchanin, Gorodetskii, and
  Il'chenko]{vyatchanin_tunable_1992}
S.~P. Vyatchanin, M.~L. Gorodetskii, and V.~S. Il'chenko.
\newblock Tunable narrow-band optical filters with modes of the whispering
  gallery type.
\newblock \emph{J. Appl. Spectrosc.}, 56\penalty0 (2):\penalty0 182--187,
  February 1992.
\newblock ISSN 0021-9037, 1573-8647.
\newblock \doi{10.1007/BF00662275}.
\newblock URL \url{http://link.springer.com/article/10.1007/BF00662275}.

\bibitem[Weigel et~al.(2012)Weigel, Esen, Schweiger, and Ostendorf]{weigel12}
T.~Weigel, C.~Esen, G.~Schweiger, and A.~Ostendorf.
\newblock Whispering gallery mode pressure sensing.
\newblock \emph{Proc. SPIE}, 8439:\penalty0 84390T--84390T--6, 2012.
\newblock \doi{10.1117/12.921759}.
\newblock URL \url{http://dx.doi.org/10.1117/12.921759}.

\bibitem[Zamora et~al.(2007)Zamora, Díez, Andrés, and
  Gimeno]{zamora_refractometric_2007}
Vanessa Zamora, Antonio Díez, Miguel~V. Andrés, and Benito Gimeno.
\newblock Refractometric sensor based on whispering-gallery modes of thin
  capillarie.
\newblock \emph{Opt. Express}, 15\penalty0 (19):\penalty0 12011--12016,
  September 2007.
\newblock \doi{10.1364/OE.15.012011}.
\newblock URL
  \url{http://www.opticsexpress.org/abstract.cfm?URI=oe-15-19-12011}.

\end{thebibliography}
\bibliographystyle{unsrt}

\end{document}